\documentclass[pra,final,twocolumn,showpacs]{revtex4-1}
\usepackage{graphicx,amsmath,amsfonts,wasysym,color}

\begin{document}


\title{Resistive and sympathetic cooling of highly-charged ion clouds in a Penning trap}

\author{M. Vogel}
\affiliation{Institut f\"ur Angewandte Physik, Technische Universit\"at Darmstadt, 64289 Darmstadt, Germany}
\author{H. H\"affner}
\affiliation{Department of Physics, University of California, Berkeley, California 94720, USA}
\author{K. Hermanspahn}
\affiliation{Institute of Environmental Science and Research, Christchurch 8041, New Zealand}
\author{S. Stahl}
\affiliation{Stahl Electronics, 67582 Mettenheim, Germany}
\author{J. Steinmann}
\affiliation{Hochschule Darmstadt, 64295 Darmstadt, Germany}
\author{W. Quint}
\affiliation{GSI Helmholtzzentrum f\"ur Schwerionenforschung, 64291 Darmstadt, Germany}

\begin{abstract}
We present measurements of resistive and sympathetic cooling of ion clouds confined in a Penning trap. For resistive cooling of a cloud consisting of one ion species, we observe a significant deviation from exponential cooling behaviour which is explained by an energy-transfer model. The observed sympathetic cooling of simultaneously confined ion species shows a quadratic dependence on the ion charge state and is hence in agreement with expectations from the physics of dilute non-neutral plasmas.
\end{abstract}

\maketitle

\section{Introduction}
Several existing and upcoming experiments with highly charged ions confined in Penning traps \cite{gho,werth} rely on effective mechanisms for cooling of the ions' motions \cite{art0,art1,spec0,spec1,kluge,fra,fra2}.
Past theoretical studies \cite{stein,gian,gian2,gorp} have investigated resistive and sympathetic cooling \cite{gho,werth,win75,wint} of highly charged ions under these conditions, but the interpretation of the sparse existing data is still subject of lively discussion. More data is required to validate simulations and assist in designing future experiments.

We have performed systematic measurements of resistive and sympathetic cooling with highly charged carbon and oxygen ions confined in a Penning trap. These were preceding steps to the measurements of the anomalous magnetic moment of the bound electron performed at the University of Mainz, Germany in collaboration with GSI, Darmstadt, Germany \cite{her,haff,haff2}, but have not been evaluated and explained so far. We discuss the results and explain them in the framework of a dedicated energy-transfer model which relates the ion-ion interactions and ion-trap interactions to the energy reservoirs and the rates of energy transfers between them.

\section{Experimental}
\subsection{Setup}
The experimental setup and the procedures have been described in detail in \cite{haff2}. Briefly, an arrangement of cylindrical Penning traps is located in the homogeneous field of a superconducting magnet and is cooled to liquid helium temperature.
\begin{figure}[h!]
\begin{center}
  \includegraphics[width=0.8\columnwidth]{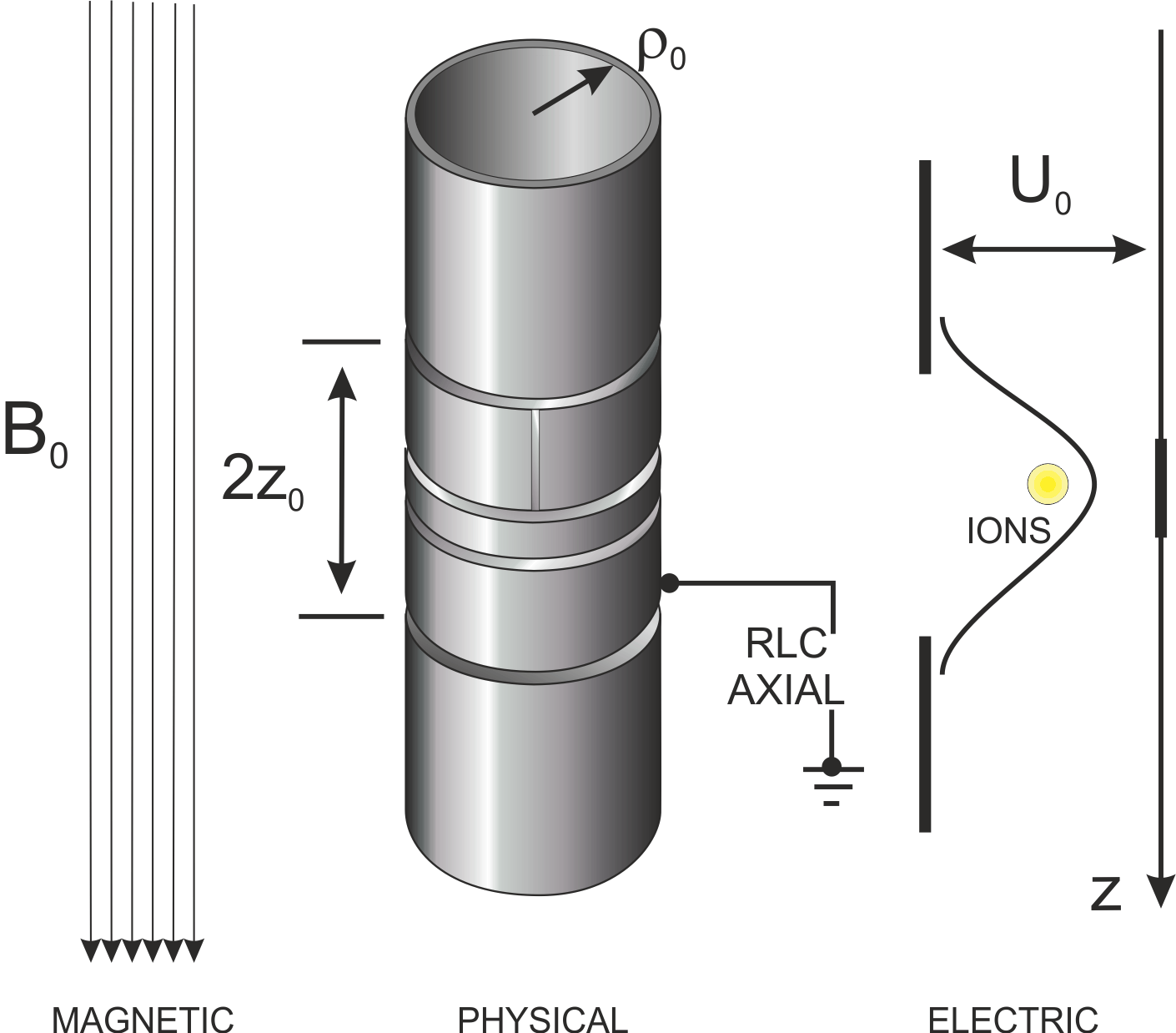}
  \caption{\small (Color online) Schematic of the Penning trap. Left: homogeneous axial magnetic field inside the trap for radial ion confinement. Center: stack of hollow cylinder electrodes forming the physical trap. Right: simplified electric potential created inside the trap for axial ion confinement.}
  \label{set}
\end{center}
\end{figure}
For the present discussion, it is sufficient to realize that ensembles of highly charged ions are produced in the cryogenic trap arrangement by electron impact ionization of atoms sputtered from a target by the same electron beam, in close similarity to the charge breeding process in electron beam ion traps \cite{joseba}. The ions are then confined for the experiments to be performed. 
In the absence of ion-ion interactions, each individual ion would perform an oscillatory motion consisting of three eigenmotions, two in the radial plane which is perpendicular to the magnetic field used for confinement, and one axial oscillation about the trap center and parallel to the experimental axis (trap axis). 
In a cloud of interacting ions, these eigenmotions are perturbed and the overall behaviour of the cloud is much more complicated. Yet, the 
axial and radial oscillatory motions can be detected non-destructively which is used to perform charge-to-mass spectrometry of the trap content. 
Pure ion clouds can be produced by resonant ejection of unwanted ions from the trap.
The number of confined ions is determined from the spectral width of a bolometric detection signal \cite{werth} and is presently of the order of a few hundreds of ions and below. Ion loss due to collision or charge exchange with neutral species has not been observed thanks to residual gas pressures below $10^{-16}$\,hPa \cite{haff2}. The detection of the ion motion, its cooling and non-destructive mass spectrometry of the trap content are performed by resonant pickup of image currents induced in trap electrodes, as will be discussed below.
 
\subsection{Ion oscillation}
In an ideal cylindrical Penning trap, a single confined ion 
obeys the axial equation of motion 
\begin{equation}
\label{eins}
\frac{\mbox{d}^2}{\mbox{d}t^2} z + \omega_z^2 z  =0,
\end{equation}
where the axial oscillation frequency $\omega_z$ follows from the axial trapping potential 
\begin{equation}
\label{harm}
V(z)=\frac{U_0C_2}{2d^2} z^2
\end{equation}
according to
\begin{equation}
\label{pott}
\omega_z^2 z = \frac{q}{m} \frac{\mbox{d}V}{\mbox{d} z},
\end{equation}
such that for the present geometry the frequency of axial oscillation $\omega_z$ is given by 
\begin{equation}
\label{z}
\omega_z=\sqrt{\frac{qU_0C_2}{md^2}} \;\;\; \mbox{with} \;\;\, d^2=\frac{z_0^2}{2}+\frac{\rho_0^2}{4}.
\end{equation} 
Here, $q$ and $m$ are the electric charge and mass of the ion, respectively, $U_0$ is the trap voltage constituting the potential well for axial confinement, $z_0$ and $\rho_0$ are the axial and radial extensions of the trap, and $C_2$ is a geometry factor which is explained in detail in \cite{bro86,gab89}.
In the present case we have $C_2 \approx 0.5412$ and for ions such as hydrogen-like carbon $^{12}$C$^{5+}$, the axial oscillation frequency $\omega_z$ is of the order of $2\pi \times 1$\,MHz. The radial oscillation frequencies $\omega_-$ (magnetron frequency) and $\omega_+$ (perturbed cyclotron frequency) are not of interest in the following, since only the axial motion is directly excited, cooled and detected in the experiment. 

When the axial trapping potential is harmonic, like the one given by equation (\ref{harm}), the axial oscillation frequency of a single ion is independent of the energy (amplitude) of this motion. If terms of orders other than $z^2$ are present, the oscillation frequency becomes energy-dependent, as has been described in detail in \cite{bro86,sens}. In real traps, this is always the case and usually efforts are undertaken to minimize these effects by appropriate choice of the trap geometry and the applied voltages \cite{gab89}. In the present case, excitation of the axial ion motion to an energy of 10\,eV per charge leads to an ion oscillation amplitude of a few mm, an average ion number density of order 10$^3$/cm$^3$ and a 10$^{-5}$ relative shift of the axial oscillation frequency, as will be discussed in detail below.

\subsection{Resistive cooling of a single ion}
\begin{figure}[h!]
\begin{center}
  \includegraphics[width=1.0\columnwidth]{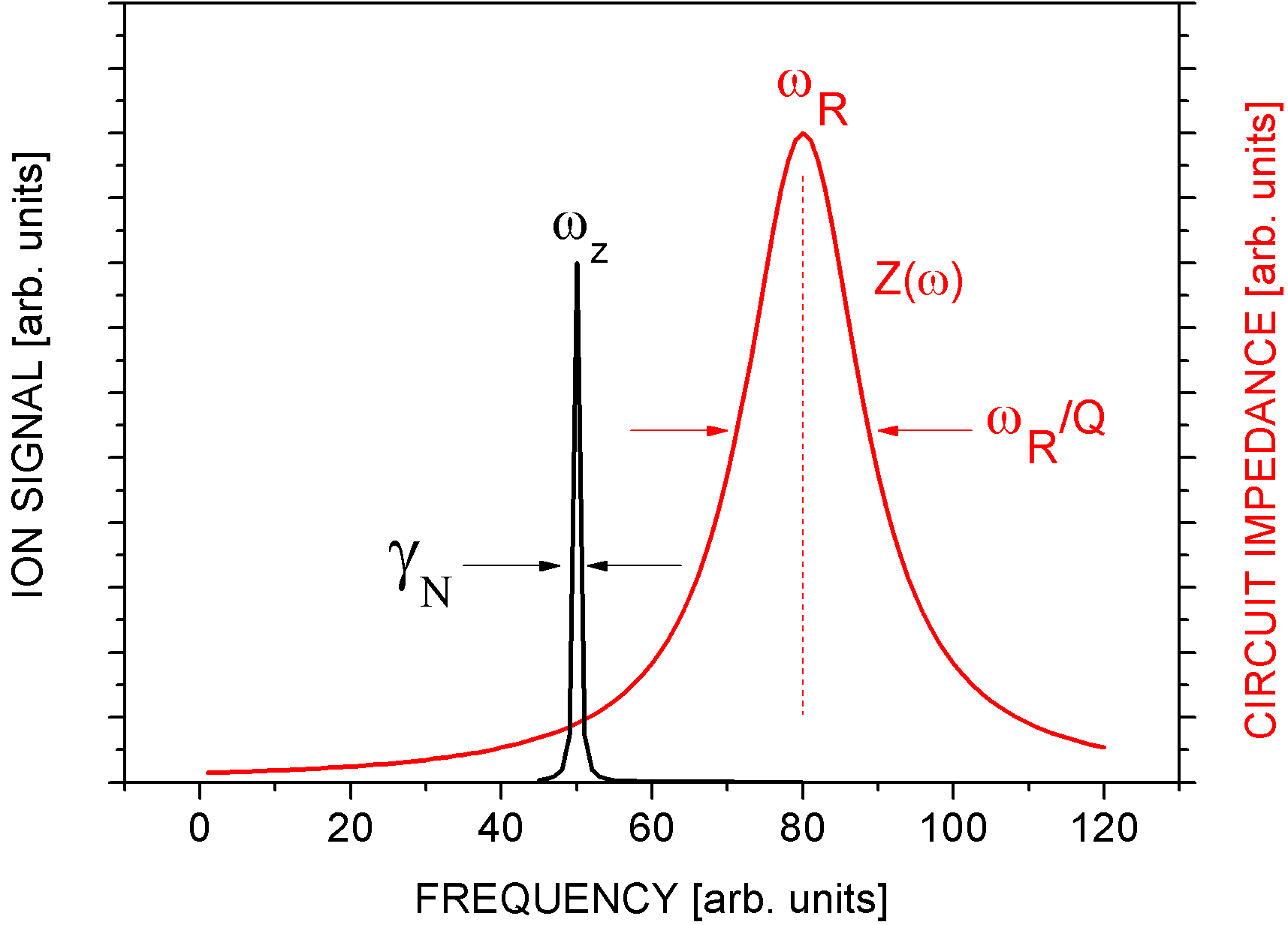}
  \caption{\small (Color online) Schematic drawing of the spectral distributions of an ion oscillation and the impedance $Z(\omega)$ of a resonant circuit. In our experiment we choose $\omega_z=\omega_R$.}
  \label{x}
\end{center}
\end{figure}

The mechanisms of resistive cooling have been explained in detail in \cite{win75,werth,gho,wint}. Briefly, an ion induces image charges in all surrounding trap electrodes \cite{shock}. When electrodes are connected by a resistance, the axial ion oscillation produces an oscillatory current through the resistance which dissipates energy from the oscillation, hence reducing the axial oscillation energy $E_z$ of the ion.
Commonly, a tuned resonance circuit with an impedance $Z(\omega_R)=R=Q\omega_RL$ is used to provide a resistance for cooling of the axial motion at $\omega_z=\omega_R$, where $Q$ (presently $Q\approx 1600$) is the quality factor and $L$ the inductance of the circuit. The quality factor $Q \gg 1$ provides large $R$ and hence efficient cooling, but limits the  range of oscillation frequencies which can be cooled to a characteristic value of $\omega_R/Q$ around $\omega_R$.
Figure \ref{x} depicts the relevant quantities, where the spectral distributions of an ion cloud and of a resonance circuit are shown.
Experimentally, we choose $\omega_z=\omega_R$ by setting the trapping voltage $U_0$ to an appropriate value. 
Note, that in general the shape of $Z(\omega)$ is a complicated function which includes the interaction with the confined ions \cite{win75,yeti}, but for the present experimental parameters we may ignore this. 

The resistive cooling may be modelled by a friction force which depends on the axial ion velocity {d$z$/d$t$}, the equation of motion then reads
\begin{equation}
\label{vier}
\frac{\mbox{d}^2}{\mbox{d}t^2} z +  \gamma_1 \frac{\mbox{d}}{\mbox{d}t}z +\omega_z^2 z  =0,
\end{equation}
where $\gamma_1$ denotes the cooling rate. 
In principle, the presence of a finite cooling rate (friction) changes the oscillation frequency according to $\omega_z'^2=\omega_z^2-\gamma_1^2/4$, but in the present situation with $\gamma_1 \ll \omega_z$ this may be neglected (see also further discussion below).
Here, as already in equations (\ref{harm}) and (\ref{pott}) we also neglect the influence of the induced image charge on the confining potential, as for a single ion the effect becomes significant only under extreme conditions \cite{voge}.
For such a weakly damped oscillator, the solution of equation (\ref{vier}) within an oscillation period $T$ is given by
\begin{equation}
\label{motio}
z=\frac{1}{\omega_z}\sqrt{ \frac{2E}{m}} \sin(\omega_z t+\phi),
\end{equation}
where E is the ion kinetic energy at the
center of oscillation.

In general, the induced current from a single ion is at any time given by
\begin{equation}
\label{i}
i_1=\frac{q}{D} \frac{\mbox{d} z}{\mbox{d} t},
\end{equation}
where the effective trap size $D$ contains all the information about the location and the geometry of the electrodes connected by the circuit with respect to the center of the ion oscillation. It is defined as $D=2z_0/\kappa$, i.e. the endcap distance $2z_0$ divided by a geometry factor $\kappa$, which has been explained in detail in \cite{bro86,gab89}. At present, with the lower correction electrode connected against a common ground (see figure \ref{set}), we have $D=5.64$\,mm. 

Following equations (\ref{motio}) and (\ref{i}), the current induced by an oscillating ion reads
\begin{equation}
\label{ii}
i_1(t)=\frac{q}{D} \frac{\mbox{d} z}{\mbox{d} t}=\frac{q}{D} \sqrt{\frac{2 E}{m}} \cos(\omega_zt+\phi).
\end{equation}
This induced current $i_1(t)$ is an oscillatory quantity of frequency $\omega_z$ and hence unqualified for an effective description over time periods as long as cooling time constants $\tau_1$. We therefore retreat to a current value time-averaged over a period of oscillation $T=2\pi/\omega_z$ and regard its evolution with advancing cooling time. We define this effective current $I_1$ by
\begin{equation}
\label{aver}
I_1^2 \equiv \langle i_1(t)^2 \rangle := \frac{1}{T} \int_0^T i_1(t)^2 \mbox{d}t,
\end{equation} 
which by use of equation (\ref{ii}) reads
\begin{equation}
\label{nine}
I_1^2 = \frac{2q^2E}{m D^2} \frac{1}{T} \int_0^T \cos^2(\omega_zt+\phi) = \frac{q^2E}{m D^2}.
\end{equation} 
The use of the kinetic energy at the center of oscillation $E$ as an energy measure requires the ion energy to remain nearly constant over a period of oscillation (weakly damped oscillator), hence the cooling rate $\gamma_1$ needs to be much smaller than the oscillation frequency, $\omega_z \gg \gamma_1$. Presently (as common for such experiments), this is well fulfilled as $2\pi \times 780\, \mbox{kHz} \gg 2\pi/(105$\,ms), where 780\,kHz is the measured axial oscillation frequency and 105\,ms is the cooling time constant $\tau_1=\gamma_1^{-1}$ of a single ion under the present conditions.

In this picture we may say that the power dissipated from the axial ion motion into the cooling circuit is given by $P_1=I_1^2 R$, hence the axial energy of the ion obeys the differential equation
\begin{equation}
\label{str}
\frac{\mbox{d}E}{\mbox{d}t}=-P_1=-I_1^2R=-\frac{q^2R}{mD^2} E = - \gamma_1E
\end{equation}
and follows an exponential decay of the kind
\begin{equation}
\label{dec}
E=E(t=0) \exp\left(-\gamma_1 t \right),
\end{equation}
where $\gamma_1$ is the single-ion cooling rate, the inverse of which is the single-ion cooling time constant 
\begin{equation}
\label{cool}
\tau_1=\gamma_1^{-1}=\frac{D^2}{R} \frac{m}{q^2},
\end{equation}
which, for the present example of $^{12}$C$^{5+}$, amounts to $\tau_1=105$\,ms at $R=9.4$\,M$\Omega$. Note, that equation (\ref{i}) assumes that the induced charge difference between the electrodes connected by the resonant circuit depends linearly on the axial coordinate of the ion. This, however, is not necessarily always the case, and higher-order (odd) terms may arise, leading to a current also at odd harmonics of the axial oscillation frequency. This has been discussed in detail in \cite{win75,wint}, but may be ignored for the present geometry. 

We also note that strictly, the axial energy $E_z$ even of a single ion undergoes fluctuations on the time scale of the cooling time constant due to the coupling to the thermal heat bath of the resistor and its electronic noise temperature. This, however may be ignored when looking at ion excitation energies of several eV as compared to the heat bath at a temperature of few Kelvin.

\subsection{Aspects of ion cloud cooling}
\label{disc}
When ensembles of ions are considered, the situation is more complicated due to the larger total number of degrees of freedom, charge effects and the presence of ion-ion interaction, hence, for larger numbers of ions, sophisticated simulation methods have to be implemented \cite{stein,gian,gian2,gorp}.

\subsubsection{From 1 to N particles in an ideal trap}
We first take a look at the extension of the above equations for $N$ ions. In full similarity to equation (\ref{aver}), for the effective current induced by $N$ ions we write
\begin{equation}
\label{strom}
I_N^2=\left\langle \left( \sum_{k=0}^N i_k(t) \right)^2 \right\rangle
= \sum_{k=1}^N \langle i^2_k \rangle + \sum_{{k,l=1}\atop{k \neq l}}^N \langle i_k i_l \rangle,
\end{equation}
where the first term on the right hand side is in analogy to equation (\ref{nine}) given by
\begin{equation}
\label{vt}
\sum_{k=1}^N \langle i^2_k \rangle = \frac{q^2 \langle 2E \rangle}{m D^2}.
\end{equation}
The quantity $\langle 2E \rangle$ (two times the mean kinetic energy) represents the total (kinetic plus potential) energy of $N$ ions. The second term on the right hand side of equation (\ref{strom}) is zero for reasons of symmetry when $N$ is sufficiently large and the phases are distributed randomly.
In this case, in analogy to equation (\ref{str}) we write the differential equation
\begin{equation}
\label{str2}
\frac{\mbox{d}\langle 2E \rangle}{\mbox{d}t}=-I_N^2R=-\frac{q^2R}{mD^2} \langle 2E \rangle = - \gamma_N \langle 2E \rangle,
\end{equation}
and find the $N$-particle cooling rate $\gamma_N$ identical to the single-particle cooling rate $\gamma_1$ in equation (\ref{cool}). Hence, the effective energy of an uncorrelated ensemble of $N$ independent ions is cooled at the same rate as a single ion. The same result has been obtained from similar arguments in \cite{gho} and \cite{ita}. A detailed derivation and discussion can also be found in \cite{werth}.

If, however, we assume $N$ identical ions which oscillate with the same amplitude and phase, in equation (\ref{strom}) we have $i_k =i_l$ and hence we find
\begin{equation}
\sum_{{k,l=1}\atop{k \neq l}}^N \langle i_k i_l \rangle = (N-1) \sum_{k=1}^N \langle i^2_k \rangle, 
\end{equation}
such that the effective current is given by
\begin{equation}
I_N^2= \sum_{k=1}^N \langle i^2_k \rangle + (N-1) \sum_{k=1}^N \langle i^2_k \rangle = \frac{Nq^2 \langle 2E \rangle}{m D^2}. 
\end{equation}
This effective current is larger than in the uncorrelated case in equation (\ref{vt}) by a factor of $N$, such that this correlated motion of $N$ ions is cooled faster than in the uncorrelated case by a factor of $N$. 

In summary, we find $\gamma_N=\gamma_1=q^2R/(mD^2)$ as the cooling rate of the mean $N$-ion energy in case of an uncorrelated motion of $N$ independent ions. If, however, the ions are correlated by moving with the same phase and amplitude, we find the corresponding cooling rate by $\gamma_N=N\gamma_1=Nq^2R/(mD^2)$.

In a perfect trap, we can separate the center-of-charge motion from motions relative to it at any given time and find the center of charge oscillating with frequency $\omega_z$. 
This requires the absence of trapping field imperfections including image charge effects on the confining potential.
For simplicity assuming only one ion species, the center of charge and the center of mass are identical and may be represented by a single particle with mass and charge of $N$ particles. Under these idealized conditions, we may use equation (\ref{vier}) for the center-of-charge axial coordinate of $N$ ions with
the substitutions $\omega'_z=\omega_z$, $V'(z)=V(z)$, $q'=Nq$, $m'=Nm$, and $\gamma_N$ instead of $\gamma_1$ is the cooling rate of the $N$-particle center of charge axial motion. As discussed above, when all ions move in phase, the cooling rate $\gamma_N$ of the axial center-of-charge motion is given by $N\gamma_1$. If the ions are completely uncorrelated, the  cooling rate is $\gamma_1$.

\subsubsection{$N$ particles in a real trap}
In reality, the potential $V'(z)$ has to include the effect of all induced image charges on the motions of all ions. In a first-order approximation the effect of image charges on the axial center-of-charge motion can be expressed as a shift of the axial oscillation frequency \cite{yeti}, but here we will neglect this effect as for the present low ion number densities and experimental resolutions it is not visible.

From Newton's third law it follows that the motion of the center of charge of an ion cloud in a perfect trap remains unaffected when ion-ion interaction is taken into account \cite{win75}. Hence, the presence of a finite space charge density will not change the motion of the center of charge of the ion cloud. 
Imperfections of the confining fields, however, will create a distribution of axial oscillation frequencies which is not intrinsic to the ion cloud. While the center of charge is well-defined at any time, its oscillation frequency spectrum does not always contain a single value under these conditions. The validity of the center-of-charge picture then depends on the experimental details, such as the actual deviation from a single axial oscillation frequency value due to trapping field imperfections.

The finite width of the axial oscillation frequency distribution in a real trap also requires to take the complex nature of the impedance $Z(\omega)$ of the resonance circuit seriously. Particularly for high charge densities, axial frequencies may differ significantly from the resonance frequency $\omega_R$ of the circuit for which it does not act as an Ohmic resistance, but creates retardation effects. At present, we may ignore this, since the trap imperfections are small and even for very high excitation energies such as 10\,eV the width of the axial frequency distribution is small when compared to the width of the resonant circuit used for cooling, as will be discussed below. 

The presence of a finite axial oscillation frequency distribution, however, provides a mechanism for the transfer of energy between axial motions and the axial center-of-charge motion. The inverse of the axial frequency width is the average rate at which axial motions transfer energy from relative motions into the axial center-of-charge motion, as has been illustrated in \cite{win75}. 
The actual observed time constant for the cooling of axial motions (which we denote by $\tau_{A}$) is hence given by the details of the interaction amongst all ions and the imperfections of the confining fields. These effects will be discussed in the following sections. The transfer mechanism between radial and axial motions will be discussed in \ref{rad}. In section \ref{mod} we will then use the results to form an energy transfer model which tries to explain the observable cooling behaviour in section \ref{cur}.

\subsubsection{Axial frequency distribution due to trapping field imperfections}
\label{axtrans}
Generally, imperfections of the confining fields will make the axial oscillation frequency of any ion dependent on its kinetic energy. Hence, for a distribution of ion energies we expect a distribution of axial oscillation frequencies. As this is an effect of the interaction between ions and the confining fields, Newton's third law does not cancel the effect on the axial center of charge. 
This is potentially an issue as the spectral width of a resonant RLC-circuit is limited by the characteristic width $\omega_R/Q$. 

The most relevant imperfections of the confining fields are deviations of the magnetic field from the homogeneous case and deviations of the electric field from the quadrupolar case. These effects have been carefully discussed in \cite{bro86,gab89,sens}. In traps like the present one, the dominant contributions to an energy-dependent shift of the axial frequency come from higher-order dependences of the axial trapping potential on the axial and radial coordinates (including mixed terms), measured by the coefficients $C_4$ and $C_6$ as defined in \cite{bro86,gab89}. Following the discussions in \cite{bro86,sens}, there is a relative shift $\Delta \omega_z/{\omega_z}$ of the axial oscillation frequency of any ion as a function of its energies $E_+$, $E_z$ and $E_-$ in the perturbed cyclotron, axial, and magnetron motion, respectively. This shift can be written as
\begin{equation}
\label{kap}
\frac{\Delta \omega_z}{\omega_z}=\frac{\omega_z^2}{2\omega_+^2} \kappa( E_+) + \frac{1}{4} \kappa(E_z) + \kappa (E_-),
\end{equation}
where $\kappa(E)$ (with $E=E_+, E_z\, \mbox{or}\, E_-$, respectively) is given by
\begin{equation}
\label{kap2}
\kappa(E) = \frac{3}{2} \frac{C_4}{C_2} \frac{ E}{qU_0} + \frac{15}{4} \frac{C_6}{C_2} \left( \frac{E}{qU_0} \right)^2.
\end{equation}
The first term in equation (\ref{kap}) may be neglected in the following, as $\omega_z^2/\omega_+^2 \ll 1$.

In the present experiment, the leading contribution to magnetic imperfection is the presence of a quadratic component $B_2z^2$ of the magnetic field. In similarity to equation (\ref{kap}) one finds \cite{sens} 
\begin{equation}
\frac{\Delta \omega_z}{\omega_z} = \frac{1}{m \omega_z^2} \frac{B_2}{B_0} E_+ + 0 E_z - \frac{1}{m \omega_z^2} \frac{B_2}{B_0} E_-,
\end{equation}
such that for the present parameters this effect is negligible when compared to the effect of electric imperfections.

Looking at these energy-dependent frequency shifts, we realize that a distribution of axial or radial kinetic energies in an ion cloud will lead to a corresponding distribution of axial oscillation frequencies within that cloud. For a thermalized ion cloud at temperature $T$, the distribution of axial and radial energies is Boltzmann-like in all motional degrees of freedom and given by
\begin{equation}
p(E)=\frac{1}{k_BT} \exp \left( -\frac{E}{k_BT} \right) \mbox{d}E,
\end{equation}
all with the expectation value $\langle E \rangle = k_BT/2$ and a typical width of the distribution of roughly $2k_BT$.
Hence, we expect the width of the axial oscillation frequency distribution over the cloud to be given by 
\begin{equation}
\label{axsh}
\sigma_z^{(T)} \approx \omega_z \left(  \frac{1}{4} \kappa(2k_BT) + \kappa (2k_BT) \right).
\end{equation}
Here, the first term is due to the axial motion and the and second term is due to the magnetron motion, and $\kappa(E)$ is again given by equation (\ref{kap2}), such that we find
\begin{equation}
\label{wid}
\sigma_z^{(T)} \approx \omega_z \left( \frac{15}{4} \frac{C_4}{C_2} \frac{k_BT}{qU_0} + \frac{75}{4} \frac{C_6}{C_2} \left( \frac{k_BT}{qU_0} \right)^2 \right).
\end{equation}
In the present experiment, we have $U_0\approx 50$\,V, $B_0=3.785$\,T, $B_2 \approx 10\mu$T/mm$^2$, $C_4$ and $C_6$ have been tuned out to approximately 10$^{-6}$ and 10$^{-3}$, respectively \cite{hafdoc}. For initial ion temperatures corresponding to 10\,eV we expect a relative width of the axial frequency distribution of around 10$^{-5}$ (still about two orders of magnitude smaller than the width of the resonant circuit) which will decrease during cooling as the ion oscillation amplitudes become smaller and less subjected to field imperfections. Note, that the end point of the cooling is around 4\,K, which corresponds to about $10^{-4}$\,eV.

\subsubsection{Collisional Thermalization}
\label{rad}
Assuming an ion cloud of arbitrary initial energy distribution and in the absence of external forces, ion-ion interactions (Coulomb collisions) thermalize the ions, eventually leading to the same Boltzmann distribution of energies within each degree of freedom. 
To quantify the time scale for this, we use the thermalization time constant ('Spitzer self-collision time'), estimated by \cite{plas}
\begin{equation}
\label{dav}
\tau_T \approx (4\pi\epsilon_0)^2\, \frac{3\sqrt{m}\, (k_BT)^{3/2}}{4 \sqrt{\pi}\, n\,q^4 \ln \Lambda},
\end{equation}
where $\ln \Lambda$ is the so-called 'Coulomb logarithm' which represents the ratio of the maximum to the minimum collision parameter possible under the given conditions, i.e. it represents the cumulative effects of all Coulomb collisions. In case of collisions amongst identical ions it is given by \cite{plas}
\begin{equation}
\ln \Lambda =  23- \ln \left(\frac{2nq^4}{e^4T^3} \right)^{1/2},
\end{equation}
where $n$ is given in units of cm$^{-3}$ and $T$ is given in units of eV.
For the present parameters, $\ln \Lambda$ is about 20, and the time constant $\tau_T$ is of the order of seconds when assuming $^{12}$C$^{5+}$ ions excited to about 10\,eV and densities of order $10^3$/cm$^3$ as discussed above. During cooling, as the density increases, the thermalization becomes more efficient. 

In particular for the present small ion numbers, however, one needs to keep in mind that ion-ion interaction at kinetic energies far from zero leads to ion number densities which fluctuate to the extent to which the collision processes are random, i.e. the density depends on both position and time, and effects which arise from charge densities will show corresponding time-dependences. Hence, all the discussion below will hold only approximately and for time-averages over an oscillation period or longer. 

\subsubsection{Energy transfer model}
\label{mod}
\paragraph{Axial motions}
As discussed above, the inverse of the axial frequency width resulting from equation (\ref{axsh}) is the average rate at which axial motions transfer energy from relative motions into the center-of-charge motion, see also the discussion in \cite{win75}.
Hence, looking at equation (\ref{wid}), for the parameters discussed above we estimate the actual cooling time constant
of axial motions to be given by
\begin{equation}
\label{coola}
\tau_A^{-1} \approx \sigma_z^{(T)}
\end{equation}
and obtain a value of several 100\,ms for $\tau_A$. As this value depends on the ion temperature, it will increase during the cooling process, hence slowing down the cooling as a function of time. Therefore, the expected cooling is not purely exponential, but an exponential $\exp(-t/\tau_A)$ with an increasing time constant $\tau_A(t)$. 

Since the spectral width of the axial oscillations determines the energy transfer to the center of charge and hence the cooling, it may be advantageous to artificially introduce field imperfections, e.g. by detuning the trap, during the time of cooling and well within the spectral width of the resonant circuit. The width $\sigma_z$ of axial oscillation frequencies determines the quality factor of the ion cloud $Q'=\omega_z/\sigma_z$. In nearly harmonic traps, this quality factor is typically much higher than the quality factor $Q$ of the resonant circuit, such that in resonance $\omega_z=\omega_R$ we may assume the impedance $Z(\omega)$ to represent a purely Ohmic resistance $R$. At present, this is the case as $Q'\approx 100000$ and $Q\approx 1600$. When the trap is made anharmonic such that no longer $Q' \gg Q$, the ion-circuit interaction becomes more complicated and equation (\ref{str2}) is no longer valid. The behaviour of $Z(\omega)$ will then limit the meaningful values of the detuning.

\paragraph{Radial motions}
Transfer of energy into the axial center-of-charge and relative axial motions can take place also from radial motions, as the radial degrees of freedom contribute to the reservoir of kinetic energies present. The radial motions are cooled with a time constant $\tau_{R}$ given by the extent they transfer energy into axial motions which are directly or indirectly cooled, as in the present experiment there is no radial resistive cooling. In section \ref{rad} we have seen that we may estimate $\tau_R$ by the Spitzer self-collision time $\tau_T$ which at present is of the order of several seconds and decreases during cooling.

The transfer of radial energy into axial degrees of freedom may be increased by active coupling of the motions. This is possible for example by irradiation of an inhomogeneous electric field at the sum or difference frequency of the motions to be coupled ('mode coupling'). This technique has been applied in several experiments \cite{cor90,dje} and is explained in detail in \cite{cor90,kre}. Mode coupling mediates a net energy transfer from the higher-energy motion to the lower-energy motion at a rate depending on parameters such as the irradiated power. It may prove helpful particularly for ion clouds of very low density.

\begin{figure}[h!]
\begin{center}
  \includegraphics[width=\columnwidth]{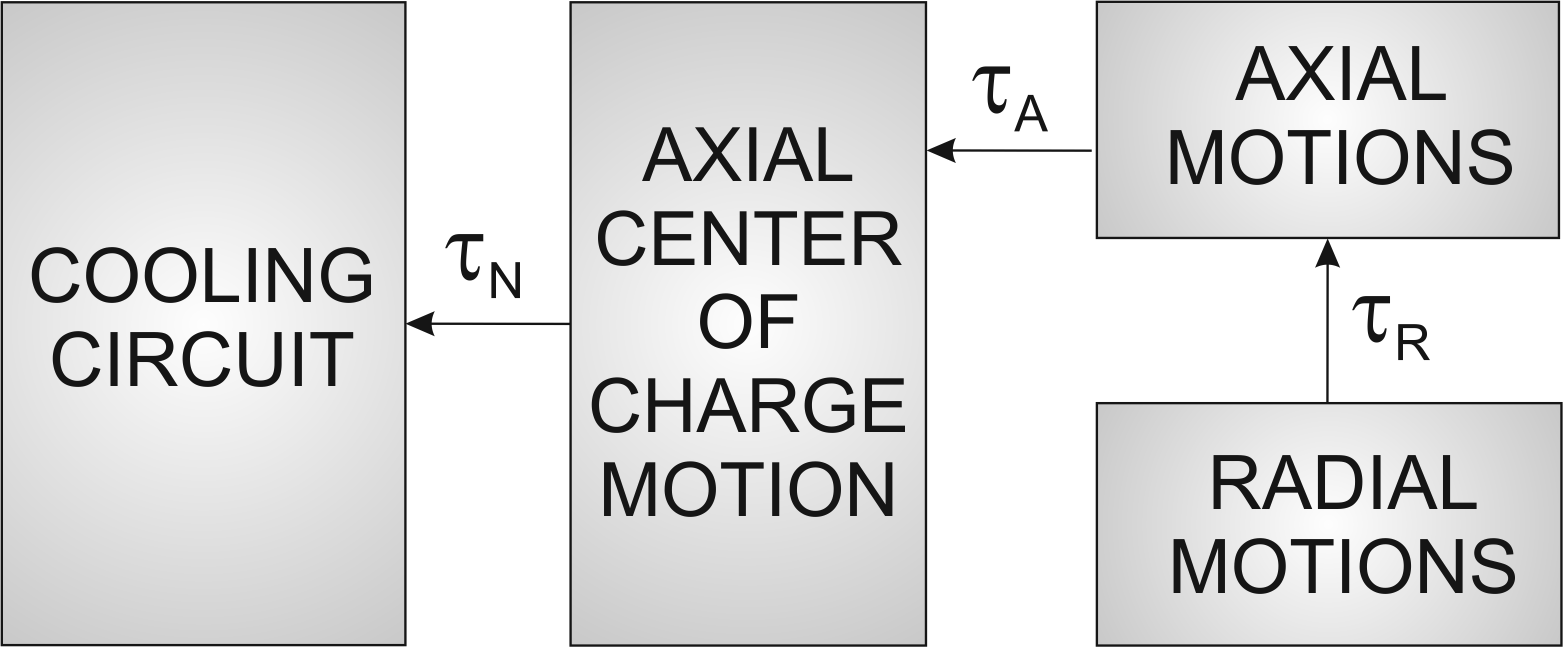}
  \caption{\small Schematic of the present energy reservoirs and energy transfers. The solid arrows indicate the direction of energy flow in the present situation, with time constants indicated.}
  \label{box}
\end{center}
\end{figure}

\paragraph{Energy reservoirs and transfer rates}
Generally, we may expect to observe three different processes: the center-of-charge cooling with its time constant $\tau_{N}$, a cooling of axial motions with a time-dependent (increasing) $\tau_{A}$ and a cooling of radial motions with a time-dependent (decreasing) $\tau_{R}$. This is a valid picture so long as the oscillation of the center of charge is well within the frequency spectrum of the resonant circuit, i.e. as long as the trapping field imperfections are small.

It is important to realize that the observability of the different components in an experiment depends crucially on the amounts of energy present in the respective motions at $t=0$. If the ion cloud is assumed to be completely thermalized, and as long as ion-ion collision rates are small enough to allow a separation of modes of oscillation, we expect
\begin{eqnarray}
 E^{(Z)}_{cc} &=& \frac{1}{3N} \cdot E, \nonumber\\
 E^{(Z)}&=& \frac{N-1}{3N} \cdot E, \label{thi} \\
 E^{(R)}&=& \frac{2N}{3N} \cdot E \nonumber
\end{eqnarray}
in the axial center-of-charge motion, the relative axial motions, and  in the radial motions, respectively, when $E$ is the total energy present. Thus, we have a hierarchy $E^{(Z)}_{cc} \ll E^{(Z)} < E^{(R)}$ with negligible relative center-of-charge energy for $N \gg 1$.
As discussed above, in such a thermalized situation, we expect cooling of the axial center-of-charge motion to occur with the single-ion cooling time constant $\tau_1$, see the discussion in section \ref{disc}. Also, as the energy transfers occur simultaneously, we expect to observe only the combined action on the ion cloud, which may not be well-separated (in the time domain) if the time constants are of similar order. Moreover, as the axial time constant $\tau_A$ is expected to increase with time, while the radial time constant $\tau_R$ is expected to decrease with time, the expected ion signal may not allow a distinction of individual contributions.

\subsubsection{Expected ion signal: Cooling curves}
\label{cur}
We have used the above model to calculate the expected ion signal ($I_N^2$ through the circuit) as a function of time for the present experimental parameters. We have assumed the initial energies $E^{(Z)}_{cc}$, $E^{(Z)}$ and $E^{(R)}$ to be thermally distributed according to equations (\ref{thi}), and the time constants $\tau_N$, $\tau_A$ and $\tau_R=\tau_T$ in the model depicted in figure \ref{box} to be given by equations (\ref{str2}), (\ref{coola}) and (\ref{dav}),
\begin{figure}[h!]
\begin{center}
  \includegraphics[width=1.0\columnwidth]{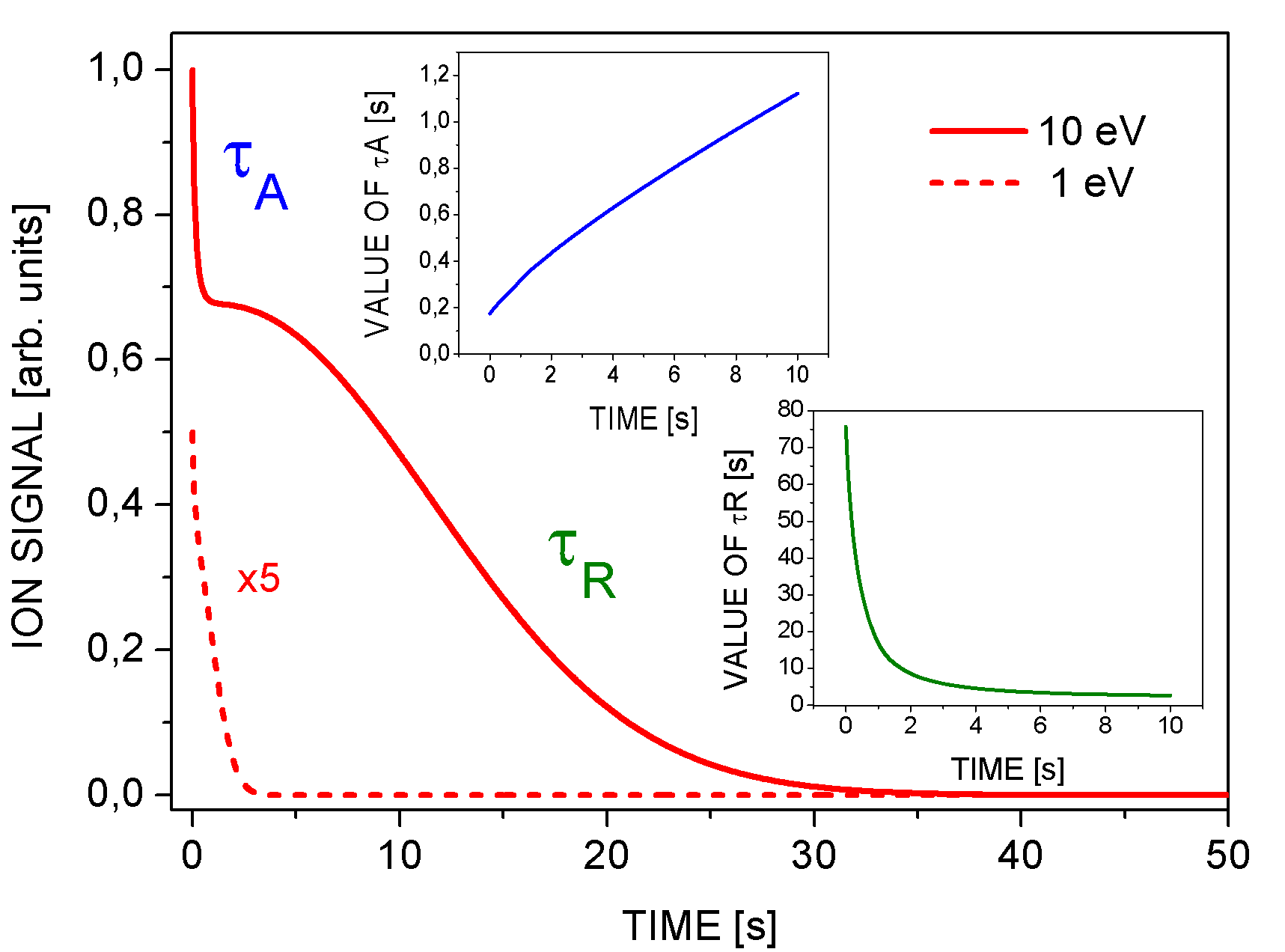}
  \caption{\small (Color online) Calculated ion signal as a function of time. Solid curve: cooling after initial excitation to 10\,eV, dotted curve: same for 1\,eV. Upper inset: value of indicated $\tau_A$ as a function of time. Lower inset: same for $\tau_R$.}
  \label{simul}
\end{center}
\end{figure}
respectively. Presently, we expect the hierarchy $\tau_N < \tau_A < \tau_R$.
For the ion number density $n$ in these equations, we have used the expression
\begin{equation}
n \approx N \left( \frac{4}{3} \pi a^3 \right)^{-1}
\end{equation}
with an effective oscillation amplitude
\begin{equation}
a \approx \sqrt{\frac{E_z d^2}{q C_2 U_0}} + \sqrt[3]{\frac{N q d^2}{3 C_2 \epsilon_0 U_0}}
\end{equation}
such that for $E_z \rightarrow 0$ the density $n$ is given by the electric space charge limit of the trap, while for large energies the density follows from the oscillation amplitudes. For the resulting ion signal $I_N^2(t)$ we use equation (\ref{str2}) with an energy $\langle E \rangle \equiv E_z(t)$ given by
\begin{equation}
\label{eee}
E_z(t)=E^{(Z)}_{cc} e^{-\frac{t}{\tau_N}} +  E^{(Z)} e^{-\frac{t}{\tau_A}} + E^{(R)} e^{-\frac{t}{\tau_R}}.
\end{equation}
Note, that as $\tau_A$ and $\tau_R$ depend on the ion energy and density, they are implicitly time-dependent and hence form coupled differential equations with (\ref{eee}) which we have evaluated numerically. 

Figure \ref{simul} shows the resulting curves. 
The small energy content in the center-of-charge motion does not lead to a visible signal decaying with $\tau_N$ on this scale. 
Note, that in a thermalized ion cloud $\tau_N$ is given by $\tau_1$.
Instead, for high initial ion excitation the curve features a fast cooling with $\tau_A$ and a slow cooling with $\tau_R$ separated by a plateau. The plateau is pronounced as for high initial ion excitation the density $n$ is low and $\tau_R$ is initially very large, see the time evolution of $\tau_R$ in the lower inset of figure \ref{simul}. The upper inset in figure \ref{simul} shows the time evolution of $\tau_A$.
To compare to the case of small initial ion excitation, the dotted curve shows the calculated cooling behaviour for an initial excitation smaller by one order of magnitude. It lacks the plateau, as for small excitation, the ion number density $n$ is sufficiently high from the beginning to produce a value of $\tau_R$ which is much smaller and comparable to the value of $\tau_A$, hence there is no clear distinction.

\subsection{Experimental procedure}
We have performed two sets of experiments, one with a pure cloud of $^{12}$C$^{5+}$ ions, we will refer to these as 'resistive cooling' measurements, and one with a mixture of different ion species (one of which is $^{12}$C$^{5+}$), to which we will refer as 'sympathetic cooling' measurements.

\subsubsection{Resistive cooling}
Upon ion creation and confinement, a single species is selected by resonant ejection of all unwanted ions from the trap. The number of remaining ions is determined from a bolometric measurement. In the present case, 30 $^{12}$C$^{5+}$ ions have been confined and investigated upon. The ions are excited by white noise excitation of the axial motion with a specific voltage amplitude $V_e$ through one endcap for 5 seconds. Upon excitation, a voltage which is proportional to the square root of the power $RI^2_N(t)$ dissipated through the resonant circuit is recorded as a function of time. 

\subsubsection{Sympathetic cooling}
A distribution of ions is produced and confined in the trap. The ions are subjected to broadband excitation of axial motions (to about 1\,eV) to produce a detectable signal.
Then, C$^{5+}$ is brought into resonance with the RLC circuit by choosing $U_0=-9.85$\,V for a variable time $t$ between 0 and 140 seconds. This is direct resistive cooling of the C$^{5+}$ ions to an axial energy which depends on the cooling time. During that time, the directly cooled C$^{5+}$ species sympathetically cools all other ion species in the trap. 

At the end of the cooling time, a spectrum is taken by ramping the trap voltage $U_0$ (in this case between -15\,V and -8\,V), thus bringing the axial oscillation frequency $\omega_z$ of every ion species briefly in resonance with the tuned circuit, hence producing a $q/m$-spectrum. For constant ion number $N$, the dissipated power $RI_N^2(t) \propto E_z(t)$, hence the detected signal is a measure of the ion energy $E_z$.
When this is repeated for different times $t$, we find a time-dependent axial ion energy and may follow the cooling process $E_z(t)$.

Once a spectrum is taken, the ions are cooled back to base temperature for the process to be repeated, starting again with excitation of all ions. This is repeated 70 times over, such that the direct cooling of C$^{5+}$ and the sympathetic cooling of all other species is observed for timespans between 0 and 140 seconds in steps of 2 seconds, hence we obtain 70 sequential spectra.

\section{Results}
\subsection{Resistive cooling}
\begin{figure}[h!]
\begin{center}
  \includegraphics[width=1.0\columnwidth]{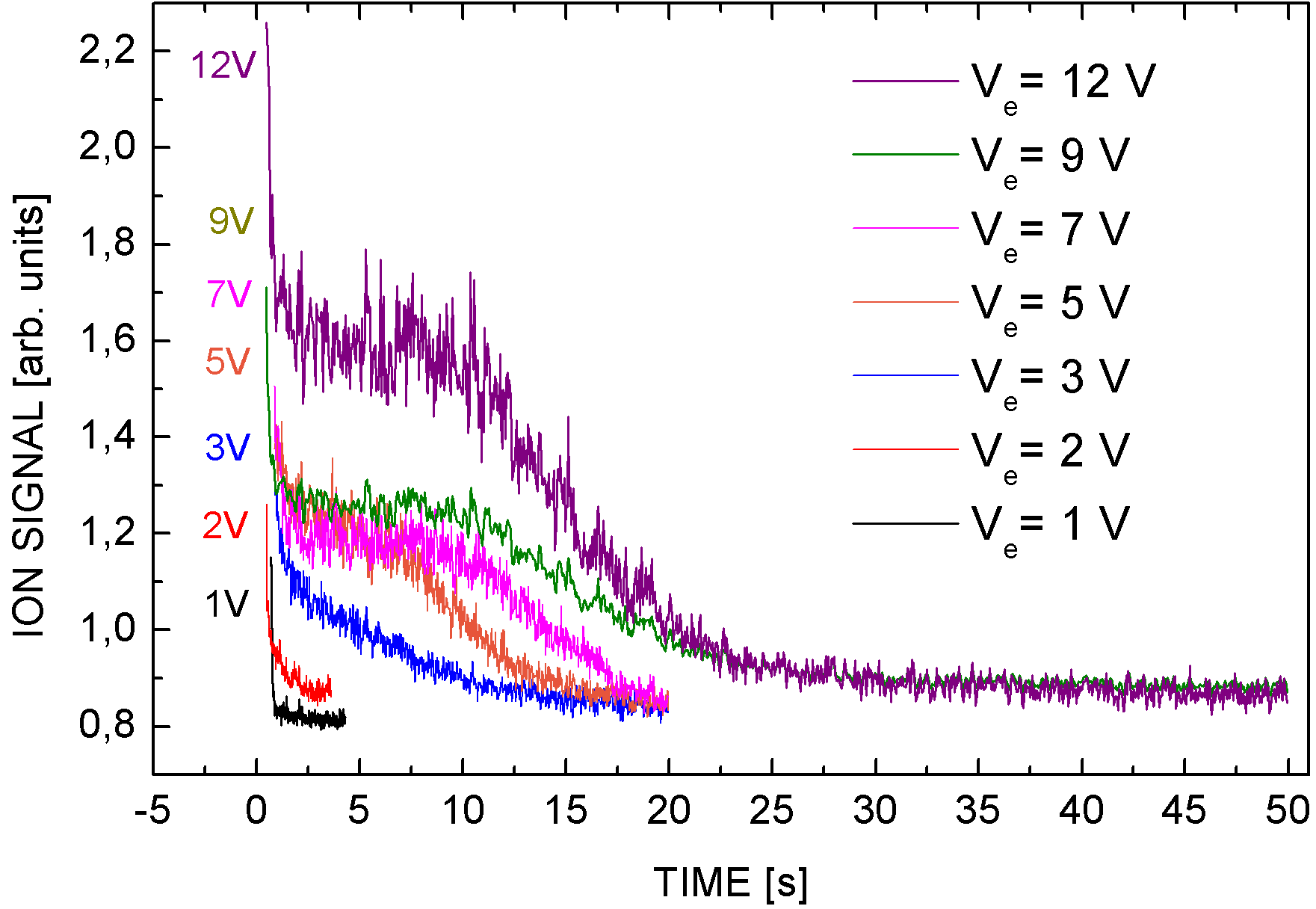}
  \caption{\small (Color online) Measured signal of 30 C$^{5+}$ ions as a function of time for different excitation amplitudes $V_e$.}
  \label{g}
\end{center}
\end{figure}

Figure \ref{g} shows the detected ion signal (squared voltage, proportional to $I_N^2$) as a function of time for different initial excitation voltage amplitudes $V_e$. 
The short-term fluctuation of the signal may be attributed to electronic noise and to fluctuations of the induced current due to ion-ion interaction.
For small excitations, the curves show a featureless decay which can be fitted with a time constant of the order of a few times the single-ion cooling time constant $\tau_1$.
From a certain excitation amplitude on (presently about 5\,V), the cooling curves show a decay with a time constant $\tau_{A}$ (of the order of a few $\tau_1$) followed by a plateau and a slow decay with a time constant $\tau_{R}$ of the order of seconds.
With increasing excitation amplitudes, this plateau becomes more pronounced, which agrees with the expected initial radial cooling being very ineffective due to the low initial ion number density $n$, see the discussion in section \ref{cur}. 
\begin{figure}[h!]
\begin{center}
  \includegraphics[width=\columnwidth]{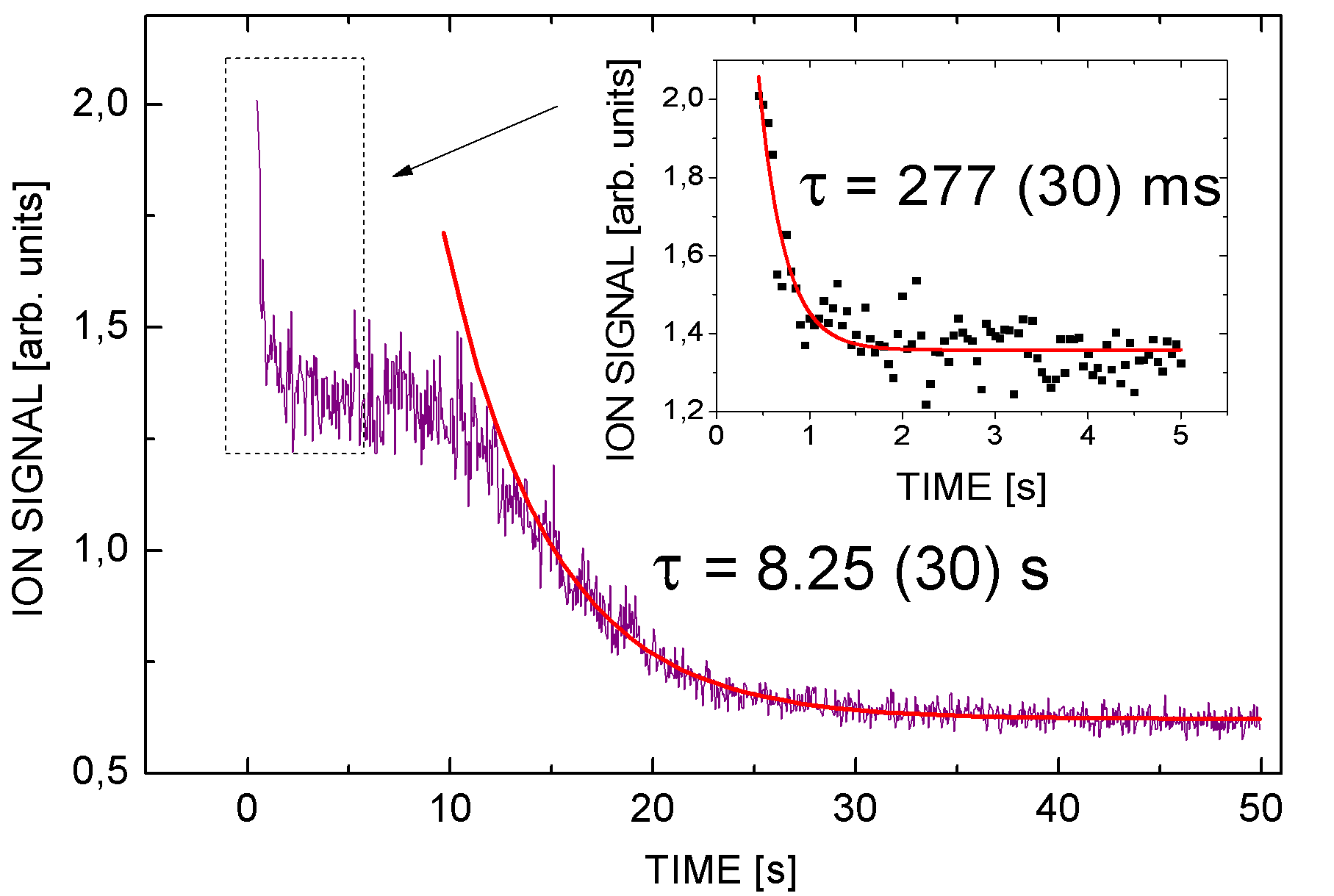}
  \caption{\small (Color online) Ion cooling curve from figure \ref{g} for the highest excitation voltage $V_e=12$\,V. The inset shows a fit to the data within the initial 5 seconds.}
  \label{c30}
\end{center}
\end{figure}
Overall, the observed cooling behaviour qualitatively agrees with the results of the model discussed in section \ref{cur}, see also figure \ref{simul}. Unfortunately, the data do not provide a solid base for a fit of the calculated curves to the data.
However, to better illustrate the two cooling domains, figure \ref{c30} shows the curve for highest initial excitation and two decays fitted to the data before and after the plateau.
For simplicity, we have neglected the time-dependence of $\tau_A$ and $\tau_R$, which is admissible when restricting the discussion to small domains far away from the plateau.

\begin{figure}[h!]
\begin{center}
  \includegraphics[width=\columnwidth]{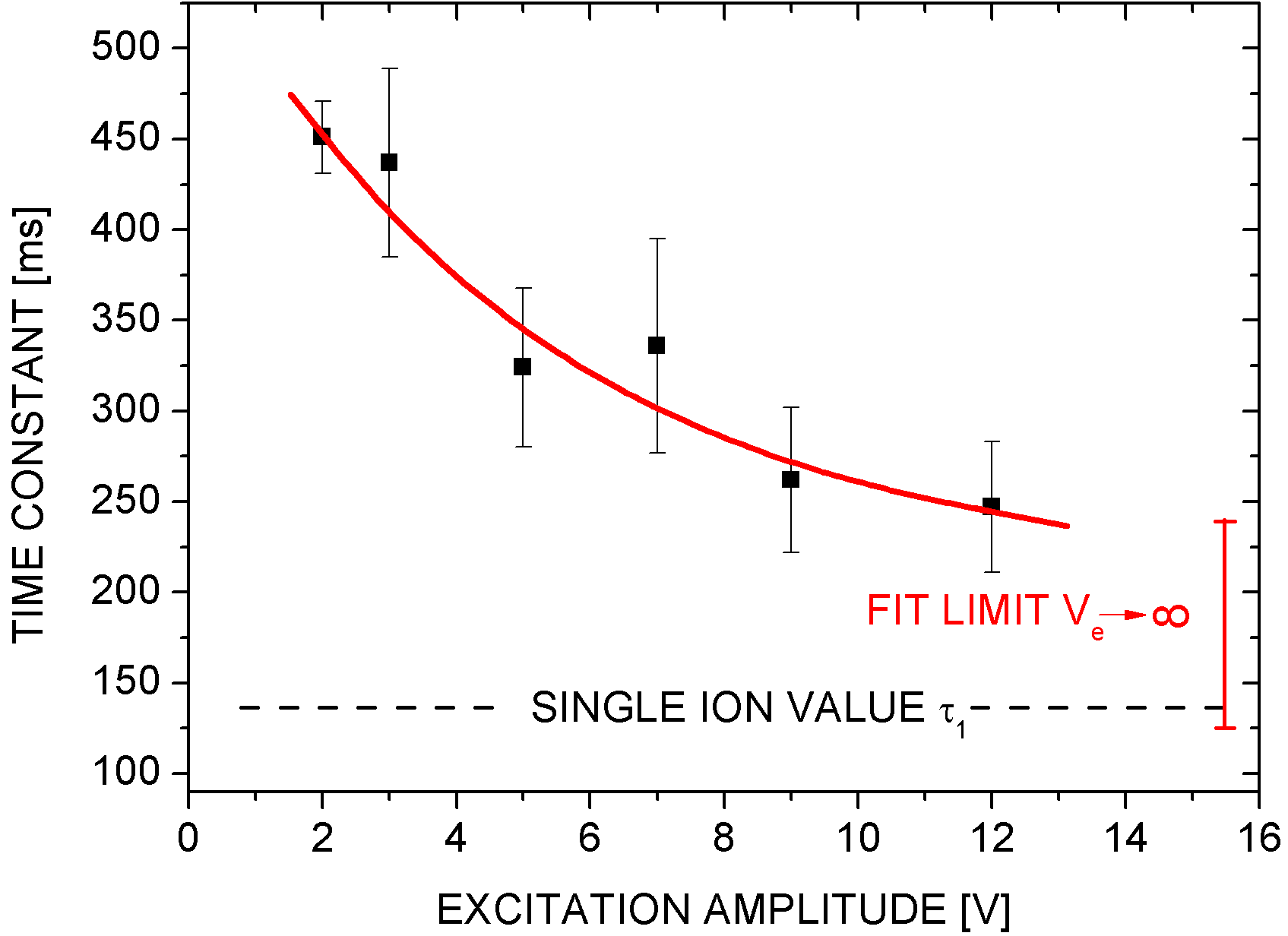}
  \caption{\small (Color online) Cooling time constants resulting from a fit to the initial 2.5 seconds of the signals as shown in figure \ref{g}.}
  \label{h}
\end{center}
\end{figure}

The initial cooling, before entering the plateau region (between $t=$0.02\,s and 2.5\,s), has been fitted separately for all curves, and results in time constants $\tau_{A}$ as plotted in figure \ref{h}. 
They tend to decrease with increasing excitation voltage $V_e$, which is in agreement with equation (\ref{coola}): for larger initial amplitudes the expected axial frequency width increases, making axial energy transfer more efficient. For large motional amplitudes, the situation approaches the case of independent ions without a fixed phase relation, for which cooling is expected to occur with the single-ion cooling time constant $\tau_1$, see the discussion in section \ref{disc}. When looking at figure \ref{h}, the data support this interpretation, as the measured single-ion cooling time constant $\tau_1=132$\,ms approximately agrees with the independent-ion limit for $V_e \rightarrow \infty$ of $(179 \pm 55)$\,ms when an exponential function is fitted to the cooling time constants $\tau_{A}$ as a function of $V_e$. The choice of an exponential is not motivated by theory and thus arbitrary, but describes the data well for all practical purposes present.

The slow component of the decay after the plateau (attributed to radial energy transfer into axial motions with a time constant $\tau_{R}$) tends to increase with increasing initial excitation: an exponential fit for $t\ge 7.5$\,s yields 4.43(13)\,s for 5\,V, 5.14(21)\,s for 7\,V, 7.11(19)\,s for 9\,V and 8.25(30)\,s for 12\,V. This agrees with the picture that the average ion number density $n$ decreases with increasing excitation, such that collisional thermalization becomes less efficient, see equation (\ref{dav}).

The cooling of the center-of-charge motion is not resolved here. For an ion motion with a common phase, the cooling would be expected to have a time constant $\tau_N=\tau_1/N$ of few milliseconds, which is beyond the current experimental time resolution. For the more realistic case of a largely thermalized motion (looking at the long initial excitation with white noise), the expected time constant is $\tau_1$ (and hence observable in this experiment), but the energy content is too small to produce a signal which allows a distinction of the time constants $\tau_1$ and $\tau_A$.  

\subsection{Sympathetic cooling}
\begin{figure}[h!]
\begin{center}
  \includegraphics[width=\columnwidth]{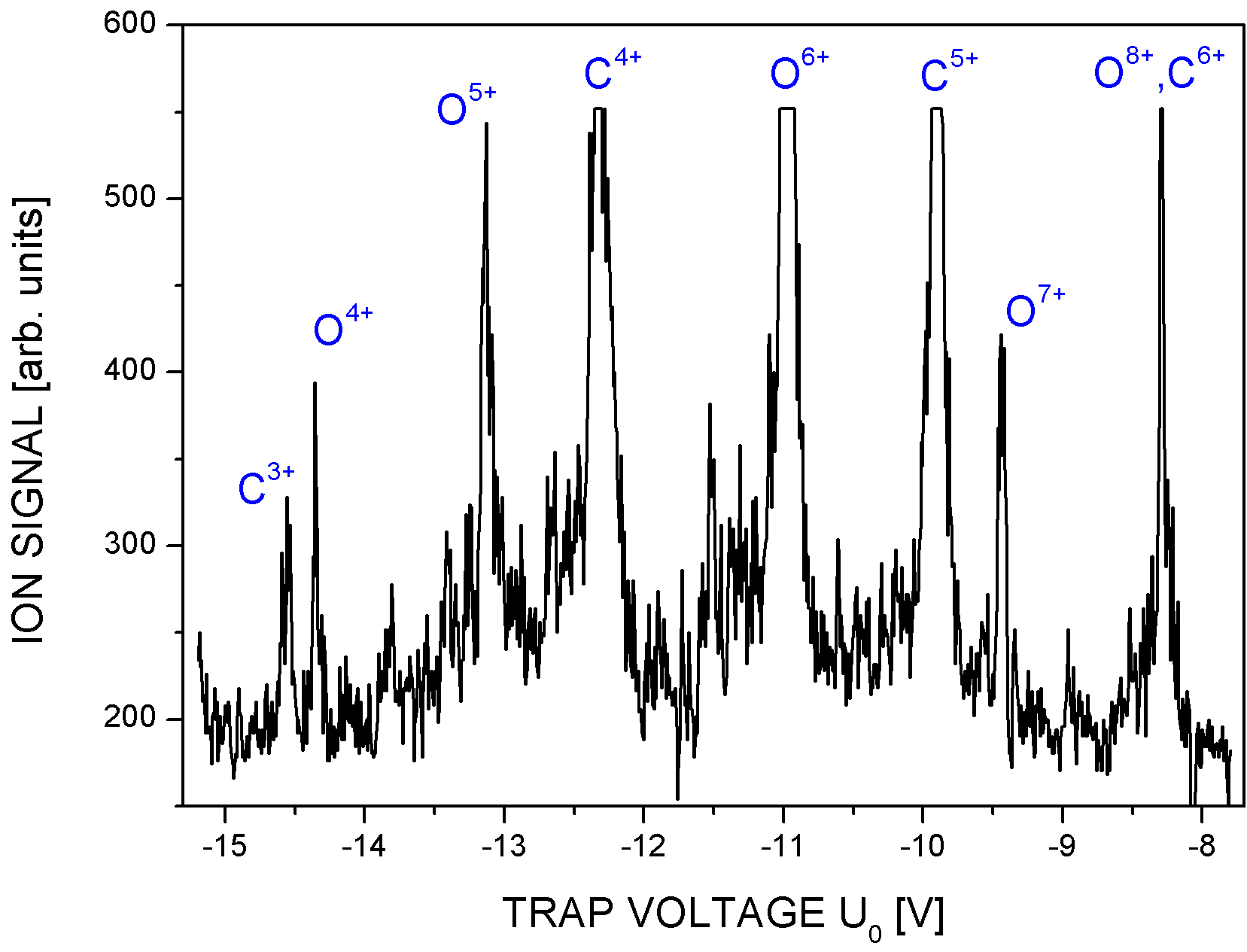}
  \caption{\small (Color online) Spectrum of the trap content upon ion creation. The square of the detection voltage $V$ is plotted as a function of the set trap voltage $U_0$.}
  \label{a}
\end{center}
\end{figure}

Figure \ref{a} shows a typical spectrum upon ion creation ($t$=0). It shows the square of the voltage signal $V$ detected across the resonant circuit as a function of the applied trap voltage $U_0$. 
The assignment of ion species to peaks is straight-forward using equation (\ref{z}) when setting $\omega_z(U_0)=\omega_R$. In this plot, the area under a peak is a measure of the ion energy times ion number, as discussed above.
After integrating over individual peaks with a simple saturation correction and background subtraction, the time evolution of the kinetic energy $E_z(t)$ of each ion species is obtained from the sequence of spectra. This is correct if no ion loss occurs during the measurement time, which has been observed to be true \cite{haff2}. 
\begin{figure}[h!]
\begin{center}
  \includegraphics[width=\columnwidth]{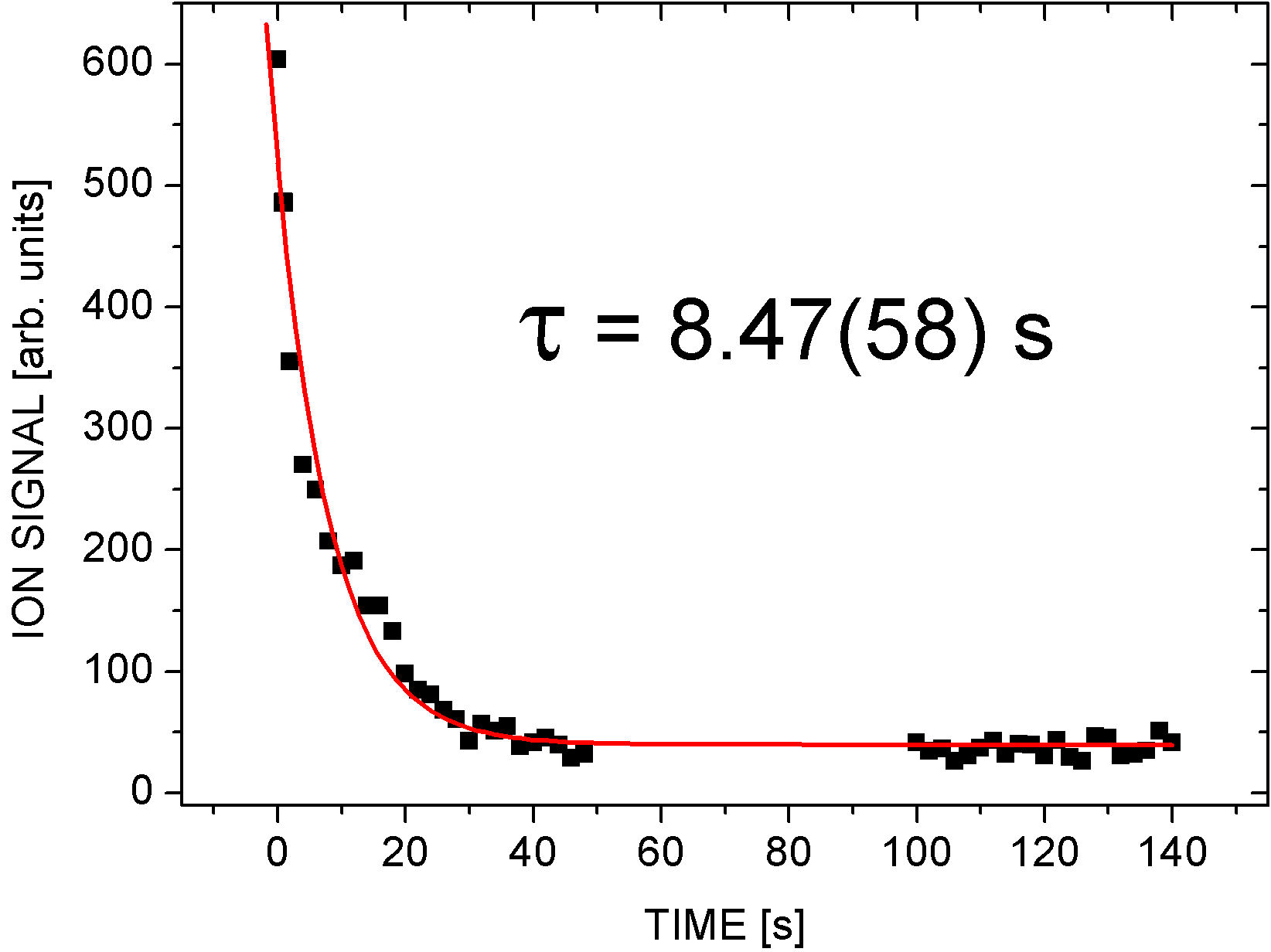}
  \caption{\small (Color online) Integrated ion signal for the C$^{4+}$ peak from figure \ref{a} as a function of the time. A simple exponential decay has been fitted to the data.}
  \label{b}
\end{center}
\end{figure}
For each ion species in the spectrum, the resulting time evolution
$E_z(t)$ can be fitted by a simple exponential decay of the kind (\ref{dec}) with different time constants for each ion species. Figure \ref{b}
shows the example of C$^{4+}$. 

We can extend the thermalization as described by equation (\ref{dav})  to two different ion species to obtain an expected time constant for sympathetic cooling. We then have to write
the Coulomb logarithm in the form \cite{plas}
\begin{equation}
\ln \Lambda = 23- \ln \left[  \frac{q_1q_2 (m_1+m_2)}{e^2 (m_1T_1+m_2T_2)}  \left(  \frac{n_1q_1^2}{e^2T_1}+\frac{n_2q_2^2}{e^2T_2}  \right)^{1/2} \right] \nonumber
\end{equation}
where again the densities $n$ are given in units of cm$^{-3}$ and the temperatures $T$ are given in units of eV. Assuming full spatial overlap of the ions, the sympathetic cooling time constant for a species '1' by a reservoir of species '2' is given by \cite{plas}
\begin{equation}
\label{elf}
\tau_S=(4 \pi \epsilon_0)^2 \frac{m_1m_2}{q_1^2q_2^2} \frac{1}{n_2 \ln \Lambda} \left( \frac{k_BT_1}{m_1} + \frac{k_BT_2}{m_2}      \right)^{3/2}.
\end{equation}
The temperature evolution of the cooled species is then given by $\partial T_1 / \partial t = (T_2-T_1)/\tau_S$. In case of several different coolants, this is generalized to $\partial T_1 / \partial t = \sum_a (T_a-T_1)/\tau^{(a)}_S$.

For the present parameters, the sympathetic cooling time constant $\tau_S$ is of the order of seconds, when $^{12}$C$^{5+}$ ions are used to cool similar ions of roughly the same or slightly smaller density.

\begin{figure}[h!]
\begin{center}
  \includegraphics[width=\columnwidth]{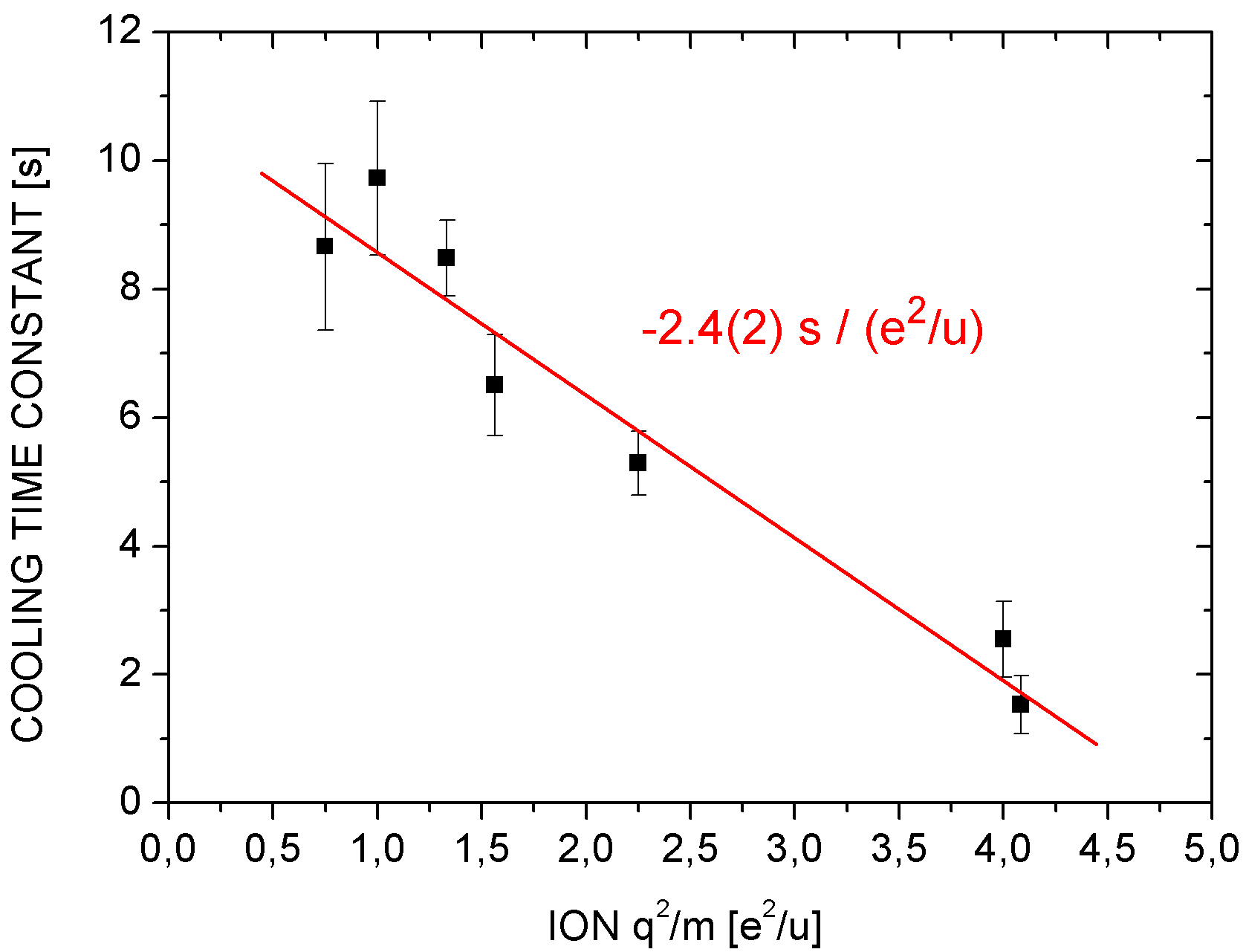}
  \caption{\small (Color online) Resulting cooling time constants for all ion species in the spectrum as a function of $q^2/m$. A linear fit has been applied to the data. }
  \label{c}
\end{center}
\end{figure}

With regard to equation (\ref{elf}), we have plotted the resulting cooling time constants as a function of the ions' squared charge divided by their mass ($q^2/m$). The result is shown in figure \ref{c}.
The cooling time constants are of order seconds, hence they agree with sympathetic cooling time constants as predicted by equation (\ref{elf}).
Obviously, a linear dependence of the cooling time constant $\tau_S$ on $q^2/m$ is supported by the data, at least for the present low ion number densities, which is also in agreement with equation (\ref{elf}): for a given coolant, the expected cooling time depends linearly on the $q^2/m$ of the ions to be cooled.
Further application of equation (\ref{elf}) to this situation is, however, not straight-forward, as some details of the interactions amongst the ions are unclear, for example whether there is centrifugal separation between species which reduces the spatial overlap etc.

Also, we are not concerned with two, but with several different species which all interact simultaneously. In principle, one would need to write down coupled cooling rate equations like (\ref{elf}) for all present species and compare to the observations, however, the present data do not allow such a detailed analysis. Still, the expected slope of the sympathetic cooling time constant with $q_1^2/m_1$ of the cooled C$^{5+}$ ions
\begin{equation}
\frac{\partial\, \tau_S}{\partial \left( \frac{q_1^2}{m_1} \right)} = -(4 \pi \epsilon_0)^2 \frac{m_2}{q_2^2} \frac{1}{n_2 \ln \Lambda} \left( \frac{k_BT_1}{m_1} + \frac{k_BT_2}{m_2}      \right)^{3/2}
\end{equation}
according to equation (\ref{elf}) of about -2.8\,s/(e$^2$/u) is in fair agreement with the measured slope in figure \ref{c} of -2.4(2)\,s/(e$^2$/u), given the unaccounted average over different ion species. 

\section{Conclusion}
We have performed measurements of resistive and sympathetic cooling of dilute clouds of highly charged ions confined in a Penning trap. Resistive cooling of a single ion species leads to non-exponential energy loss with a fast and a slow component, which, depending on the initial level of ion excitation, may be well-separated in time and produce visible features such as a pronounced plateau between the components. 

The faster of the observed cooling time constants $\tau_{A}$ is in fair agreement with the value expected from the present trapping field imperfections and ion-ion interactions. This observed cooling rate is given by the transfer of axial energy into the axial center-of-charge motion and is generally time-dependent, as the rate of energy transfer depends on the ion kinetic energy itself. The time constant $\tau_A$ is hence not a true constant, but increases as a function of cooling time.

The slower of the observed components has a time constant $\tau_{R}$ of the order of seconds and is attributed to energy transfer from radial motions into axial motions by Coulomb collisions. $\tau_R$ is also not a true constant, as the energy transfer rate through collisions is density- and energy-dependent. Hence, as a function of cooling time, its rate increases with the increasing ion number density and the decreasing ion energy. 

The cooling time constant $\tau_N$ of the axial center-of-charge motion has not been resolved, as for a non-thermal ion cloud the cooling is expected outside of the time scale of observation, while for a thermalized ion cloud the expected energy content of this motion is too small to be observed directly in the present experiment. 


Sympathetic cooling of a distribution of species by resistively cooled ions which are simultaneously trapped leads to motional energy loss of all confined species which can be described by exponential decays, with time constants depending roughly linearly on the squared-charge to mass ratio $q^2/m$ of the respective species. The observed sympathetic cooling time constants of the ion clouds are of the order of seconds and are in fair agreement with expectations from physics of non-neutral plasmas. This is also true for their scaling with the charge-to-mass ratios of the ions. 

More exact quantitative statements about this suffer from the fact that cooling of clouds strongly depends on details of the electronic cooling and detection scheme, and on initial conditions prior to cooling, such as the ion distributions in position and momentum space, which commonly are largely unknown in experiments. This is also a problem when simulations and experimental findings are to be compared.

To fully understand the behaviour of ion clouds under resistive and sympathetic cooling, it appears necessary to perform further systematic measurements, particularly of ion-number dependent quantities, and to make direct comparisons with simulations, which above all demands well-defined initial conditions prior to cooling.

\section{Acknowledgement} We thank R.C. Thompson (Imperial College London) for inspiring discussions and helpful comments.


\begin{thebibliography}{99}
\bibitem{gho} P. Ghosh, \textit{Ion Traps}, Oxford University Press, Oxford (1995) 
\bibitem{werth} G. Werth, V.N. Gheorghe and F.G. Major, \textit{Charged Particle Traps}, Springer, Heidelberg, 2005 
\bibitem{art0} W. Quint, D.L. Moskovkhin, V.M. Shabaev and M. Vogel, Phys. Rev. A {\bf 78} (2008) 032517
\bibitem{art1} D. von Lindenfels et al., Phys. Rev. A {\bf 87} (2013) 023412
\bibitem{spec0} M. Vogel et al. Rev. Sci. Inst. {\bf 76} (2005) 103102
\bibitem{spec1} Z. Andelkovic et al., Phys. Rev. A {\bf 87} (2013) 033423

\bibitem{kluge} H.-J. Kluge et al., Advances in Quantum Chemistry {\bf 53} (2007) 83
\bibitem{fra} F. Herfurth et al., Int. J. Mass Spectr. {\bf 251} (2006) 266
\bibitem{fra2} F. Herfurth et al., Hyp. Int. {\bf 173} (2006) 93

\bibitem{stein} J. Steinmann, J. Gro\ss, F. Herfurth and G. Zwicknagel, AIP Conf. Proc. {\bf 1521} (2013) 240

\bibitem{gian} G. Maero et al., Appl. Phys. B {\bf 107} (2012) 1087

\bibitem{gian2} G. Maero, \textit{Cooling of highly charged ions in a Penning trap for HITRAP}, PhD thesis, University of Heidelberg (2008)

\bibitem{gorp} S. van Gorp et al., Nuclear Instruments and Methods in Physics Research A {\bf 638} (2011) 192

\bibitem{win75} D.J. Wineland and H.G. Dehmelt, J. Appl. Phys. {\bf 46} (1975) 919
\bibitem{wint} D. Winters et al., J. Phys. B {\bf 39} (2006) 3131

\bibitem{her} N. Hermanspahn et al., Phys. Rev. Lett. \textbf{84} (2000) 427

\bibitem{haff} H. H\"affner, et al., Phys. Rev. Lett. {\bf 85} (2000) 5308
\bibitem{haff2} H. H\"affner et al., {Eur. Phys. J. D} {\bf 22} (2003) 163

\bibitem{joseba} J. Alonso et al., {Rev. Sci. Instr.} {\bf 77} (2006) 03A901

\bibitem{bro86} L.S. Brown and G. Gabrielse, {Rev. Mod. Phys.} {\bf 58} (1986) 233

\bibitem{gab89} G. Gabrielse, L. Haarsma and S.L. Rolston, Int. J. Mass Spectr. Ion Proc. {\bf 88} (1989) 319
\bibitem{sens} M. Vogel, W. Quint and W. N\"ortersh\"auser, Sensors {\bf 10}  (2010) 2169
\bibitem{shock} W. Shockley, J. Appl. Phys. {\bf 9} (1938) 635
\bibitem{voge} M. Vogel and W. Quint, \textit{Magnetic moment of the bound electron} in: \textit{Fundamental Physics in Particle Traps}, Springer (2014)
\bibitem{ita} W.M. Itano, J.C. Bergquist, J.J. Bollinger and D.J. Wineland,
Physica Scripta  {\bf T59} (1995) 106

\bibitem{yeti} X. Feng et al., J. Appl. Phys. {\bf 79} (1996) 8
\bibitem{prol} J.B. Jeffries, S.E. Barlow and G.H. Dunn, Int. J. Mass Spectrom. Ion Process. {\bf 54} (1983) 169
\bibitem{li} G. Li, S. Guan and A.G. Marshall, J. Am. Soc. Mass. Spectrom. {\bf 9} (1998) 473
\bibitem{hafdoc} H. H\"affner, \textit{Pr\"azisionsmessung des magnetischen Moments des Elektrons in wasserstoff\"ahnlichem Kohlenstoff}, PhD thesis, University of Mainz (2000)
\bibitem{plas} NRL PLASMA FORMULARY, J.D. Huba, Beam Physics Branch, Plasma Physics Division, Naval Research Laboratory, Washington, DC 20375 (2013)
\bibitem{cor90} E.A. Cornell, R.M. Weisskoff, K.R. Boyce, D.E. Pritchard,
Phys. Rev. A {\ 41}, 312 (1990)
\bibitem{dje} S. Djekic et al., Eur. Phys. J. D {\bf 31} (2004) 451
\bibitem{kre} M. Kretschmar, AIP Conf. Proc. {\bf 457}, 242 (1999)





\end{thebibliography}
\end{document}